\documentclass[prl,twocolumn,showpacs,preprintnumbers,amsmath,amssymb,floats]{revtex4-1}
\usepackage{graphicx}
\usepackage{amsmath}
\usepackage{epstopdf}
\usepackage{amsbsy}
\usepackage{color}
\usepackage{dcolumn}
\usepackage{bm}

\DeclareMathOperator{\tr}{Tr}

\begin{document}

\title{Unconventional disorder effects in correlated superconductors}
\author{Maria N. Gastiasoro$^1$, Fabio Bernardini$^2$, and Brian M. Andersen$^{1*}$}
\affiliation{$^1$Niels Bohr Institute, University of Copenhagen, Juliane Maries Vej 30, 2100 Copenhagen,
Denmark\\
$^2$CNR-IOM-Cagliari and Dipartimento di Fisica, Universit\`{a} di Cagliari, 09042 Monserrato, Italy}

\date{\today}

\maketitle

{\bf The understanding of disorder has profoundly influenced the development of condensed matter physics, explaining such fundamental effects as, for example, the transition from ballistic to diffusive propagation, and the presence of quantized steps in the quantum Hall effect. For superconductors, the response to disorder reveals crucial information about the internal gap symmetries of the condensate, and thereby the pairing mechanism itself. The destruction of superconductivity by disorder is traditionally described by Abrikosov-GorÕkov (AG) theory,\cite{ag1,ag2} which however ignores spatial modulations and ceases to be valid when impurities interfere, and interactions become important. Here we study the effects of disorder on unconventional superconductors in the presence of correlations, and explore a completely different disorder paradigm dominated by strong deviations from standard AG theory due to generation of local bound states and cooperative impurity behavior driven by Coulomb interactions. Specifically we explain under which circumstances magnetic disorder acts as a strong poison destroying high-T$_c$ superconductivity at the sub-1\% level, and when non-magnetic disorder, counter-intuitively, hardly affects the unconventional superconducting state while concomitantly inducing an inhomogeneous full-volume magnetic phase. Recent experimental studies of Fe-based superconductors (FeSC) have discovered that such unusual disorder behavior seem to be indeed present in those systems.}

For cuprates, heavy-fermions, and FeSC the study of disorder currently constitutes a very active line of research, motivated largely by the fact that these systems are made superconducting by "chemical disordering" (charge doping), but also boosted by controversies of the correct microscopic model, and a rapid development of local experimental probes.~\cite{balatsky,fischer,Alloul09} Focusing on multi-band FeSC, disorder studies have proven exceptionally rich and strongly material dependent.~\cite{hoffman11} Scanning tunneling spectroscopy found a plethora of exotic atomic-sized impurity-generated states,~\cite{grothe12,yang,chi14,yin15} NMR and neutrons observed clear evidence of glassy magnetic behavior,~\cite{dioguardi,XLu14} and $\mu$SR discovered magnetic phases generated by non-magnetic disorder.~\cite{sanna_prl,sanna_prb} The resulting complex inhomogeneous phases and their properties in terms of thermodynamics and transport constitute an important open problem in the field. 

Here, we present a theoretical study of correlation-driven emergent impurity behavior of both magnetic and nonmagnetic disorder in unconventional $s\pm$ multi-band superconductors. 
For the case of magnetic disorder, we find that correlations anti-screen the local moment, and significantly enhance the inter-impurity Ruderman-Kittel-Kasuya-Yosida (RKKY) exchange interactions by inducing non-local long-range magnetic order which operates as an additional competitor to superconductivity. 
This results in an aggressive $T_c$ suppression rate where superconductivity is wiped out by sub-1\% concentrations of disorder. 
This mechanism explains the "poisoning effect" discovered in Mn-substituted optimally doped (OD) LaFeAsO$_{1-x}$F$_x$ (La-1111) pnictide where less than 0.2\% Mn is enough to suppress the optimal $T_c\sim 30$K to zero, well beyond standard AG behavior.~\cite{sato10,hammerath14} 
By contrast, for non-magnetic disorder the $s\pm$ superconducting state is largely immune to disorder, in agreement with earlier one-band studies, finding that correlations enhance the screening of disorder potentials and thereby reduce pair-breaking and scattering rates compared to the non-interacting case.~\cite{andersen08,garg08,fukushima09,guzman13,vlad15} 
In the current multi-orbital case, however, additional impurity-generated bound states play an important unexpected role in supporting $T_c$. 
This resilience to non-magnetic disorder is remarkable since favorable clusters of impurities locally pin magnetic order, eventually causing a volume-full inhomogeneous magnetic state which coexists with superconductivity. 
These latter results  are in agreement with extensive systematic experimental studies of Ru-substituted 1111 superconducting materials. 

{\it Model.} Interactions are included in the model by the standard multi-orbital Hubbard term 
\begin{align}
 \label{eq:Hint}
 \mathcal{H}_{int}&=U\sum_{\mathbf{i},\mu}n_{\mathbf{i}\mu\uparrow}n_{\mathbf{i}\mu\downarrow}+(U'-\frac{J}{2})\sum_{\mathbf{i},\mu<\nu,\sigma\sigma'}n_{\mathbf{i}\mu\sigma}n_{\mathbf{i}\nu\sigma'}\\\nonumber
&\quad-2J\sum_{\mathbf{i},\mu<\nu}\vec{S}_{\mathbf{i}\mu}\cdot\vec{S}_{\mathbf{i}\nu}+J'\sum_{\mathbf{i},\mu<\nu,\sigma}c_{\mathbf{i}\mu\sigma}^{\dagger}c_{\mathbf{i}\mu\bar{\sigma}}^{\dagger}c_{\mathbf{i}\nu\bar{\sigma}}c_{\mathbf{i}\nu\sigma},
\end{align}
where $\mu,\nu$ are orbital indices, ${\mathbf{i}}$ denotes lattice sites, and $\sigma$ is the spin. 
The interaction includes intraorbital (interorbital) repulsion $U$ ($U'$), the Hund's coupling $J$, and the pair hopping energy $J'$. 
We assume $U'=U-2J$ and $J'=J$ and fix $J=U/4$. 
Non-magnetic and magnetic disorder give rise to terms of the form $\mathcal{H}_{imp}=\sum_{\mu\{\mathbf{i^*}\}}V_{\mu} n_{\mathbf{i^*}\mu}$ and $\mathcal{H}_{imp}=I\sum_{\{\mathbf{i^*}\}\mu\sigma}\sigma S_\mu c_{\mathbf{i^*}\mu\sigma}^{\dagger}c_{\mathbf{i^*}\mu\sigma}$, respectively. 
Here $V_{\mu}$ ($S_\mu$) denotes the impurity potential (magnetic moment) in orbital $\mu$ at the disorder sites given by the set $\{\mathbf{i^*}\}$ coupled to the charge (spin) density of the itinerant electrons. For concreteness we focus on FeSC and hence use a five-band model 
\begin{equation}
 \label{eq:H0}
\mathcal{H}_{0}=\sum_{\mathbf{ij},\mu\nu,\sigma}t_{\mathbf{ij}}^{\mu\nu}c_{\mathbf{i}\mu\sigma}^{\dagger}c_{\mathbf{j}\nu\sigma}-\mu_0\sum_{\mathbf{i}\mu\sigma}n_{\mathbf{i}\mu\sigma},
\end{equation}
with tight-binding parameters appropriate for 1111 pnictides~\cite{ikeda10}. 
The model $\mathcal H_0+\mathcal H_{int}$ exhibits a transition to a bulk SDW phase at a critical repulsive interaction $U_c$, and we parametrize the interactions in terms of $u=U/U_c$. Superconductivity is included by
\begin{equation}
 \mathcal{H}_{BCS}=-\sum_{\mathbf{i}\neq \mathbf{j},\mu\nu}[\Delta_{\mathbf{ij}}^{\mu\nu}c_{\mathbf{i}\mu\uparrow}^{\dagger}c_{\mathbf{j}\nu\downarrow}^{\dagger}+\mbox{H.c.}],
\end{equation}
with $\Delta_{\mathbf{ij}}^{\mu\nu}=\sum_{\alpha\beta}\Gamma_{\mu\alpha}^{\beta\nu}(\mathbf{r_{ij}})\langle\hat{c}_{\mathbf{j}\beta\downarrow}\hat{c}_{\mathbf{i}\alpha\uparrow}\rangle$ being the superconducting order parameter, and $\Gamma_{\mu\alpha}^{\beta\nu}(\mathbf{r_{ij}})$ denoting the effective pairing strength between sites (orbitals) $\mathbf{i}$ and $\mathbf{j}$ ($\mu$, $\nu$, $\alpha$ and $\beta$). In agreement with a general $s^\pm$ pairing state in FeSC, we include next-nearest neighbor (NNN) intra-orbital pairing. For further computational details and parameter dependence, we refer to the Supplementary Material (SM).

\begin{figure}[b]
\begin{center}
\includegraphics[width=0.98\columnwidth]{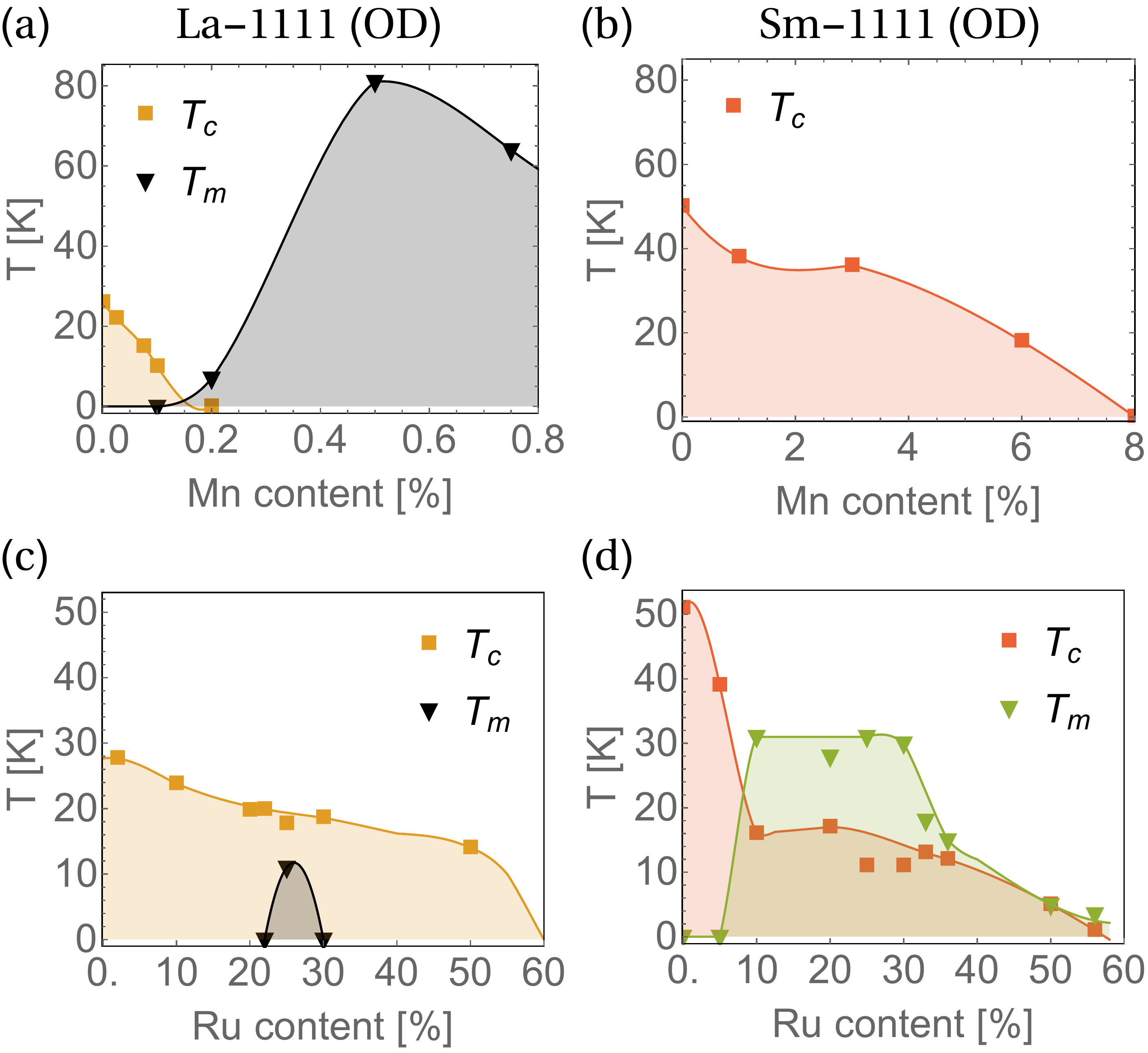}
\end{center}
\caption{Experimentally obtained superconducting $T_c$ and magnetic $T_m$ transition temperatures in OD La-1111 (a,c) and Sm-1111 (b,d) versus magnetic disorder (a,b) and non-magnetic disorder (c,d). 
The data was adapted from Refs. \onlinecite{sanna_prl,sanna_prb,sato10,hammerath14,singh11}. }
\label{fig:1}
\end{figure}

{\it Magnetic disorder.} 
The study of magnetic disorder is motivated largely by the following remarkable experimental facts shown in Figs.~\ref{fig:1}(a,b): in OD La-1111 with $T_c\sim 30$K a mere $\sim0.2\%$ magnetic Mn ions is enough to destroy the superconducting state.\cite{sato10}  This extreme destruction rate of bulk superconductivity has been recently dubbed "the poisoning effect"~\cite{hammerath14}. 
Interestingly, immediately beyond $\sim 0.2\%$, the same minute amount of Mn ions generate a static magnetic phase with full volume fraction and sizable magnetic transition temperature $T_m$. Recently it was found that this magnetic phase is $(\pi,0)$-ordered with a concomitant orthorhombic structural transition similar to the undoped system.~\cite{carretta16}
By contrast, for SmFeAsO$_{1-x}$F$_x$ (Sm-1111) near optimal doping, the corresponding $T_c$ suppression rate is much slower with $\sim 8\%$ of Mn being required to wipe out superconductivity\cite{singh11} [Fig.~\ref{fig:1}(b)]. 

\begin{figure}[]
\begin{center}
\includegraphics[width=0.9\columnwidth]{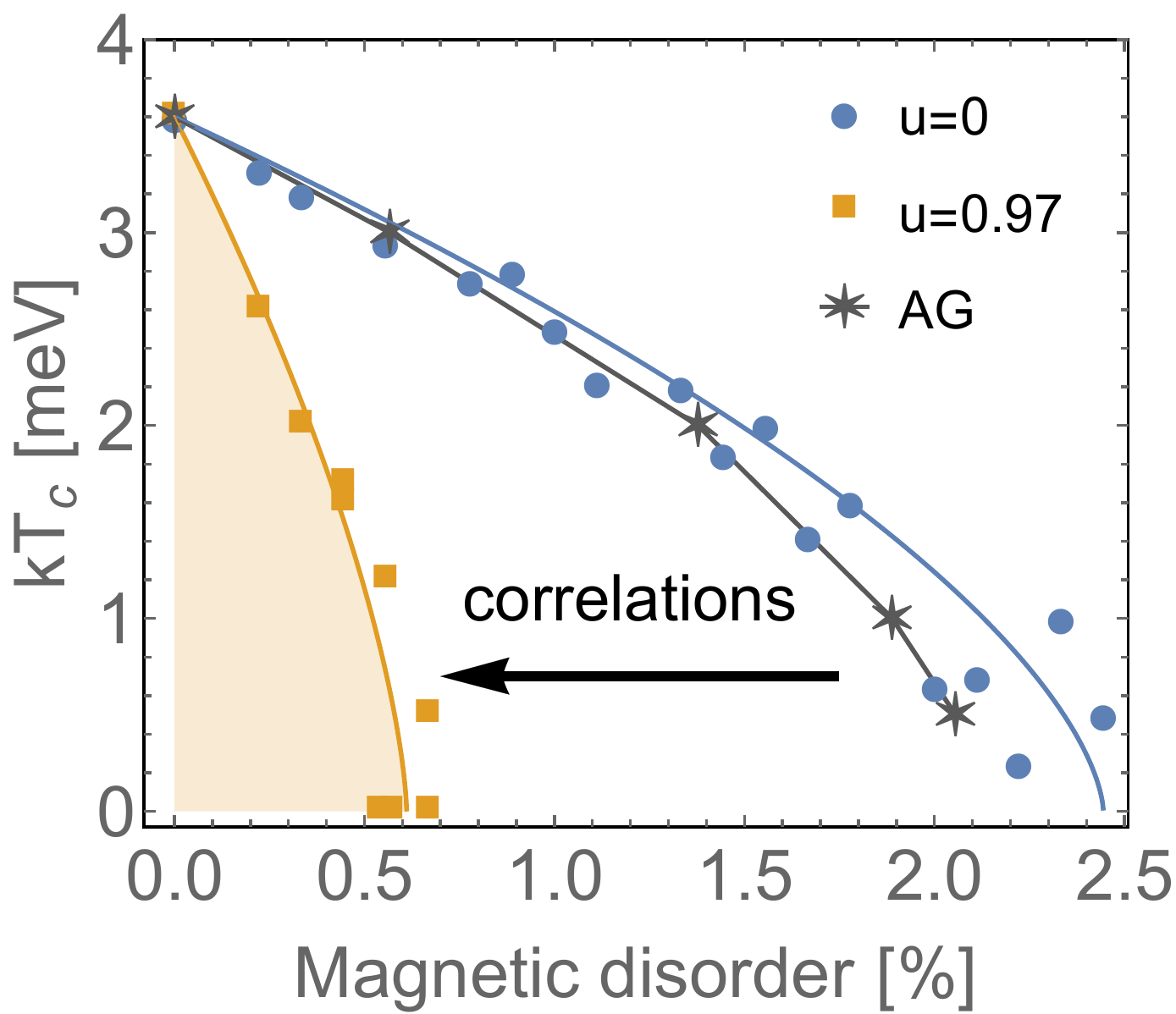}
\end{center}
\caption{{\bf Superconducting critical temperature $T_c$ as a function of magnetic impurity concentration.} 
Impurity moments are destructive for superconductivity, and the $T_c$ suppression rate is strongly modified by electronic correlations as seen by comparing the two $T_c$ curves at $u=0$ and $u=0.97$. 
The $u=0$ curve follows the behavior described by standard AG-theory. All impurity moments are modelled by orbitally independent exchange with $I S_\mu=0.38$ eV. As seen, correlations act to poison the superconducting state suppressing it entirely (at all sites) after only $\sim0.5\%$ disorder as seen by the orange curve. Larger $u$ exhibits even more severe suppression rates.
}
\label{fig:2}
\end{figure}

\begin{figure}[t]
\begin{center}
\includegraphics[width=0.9\columnwidth]{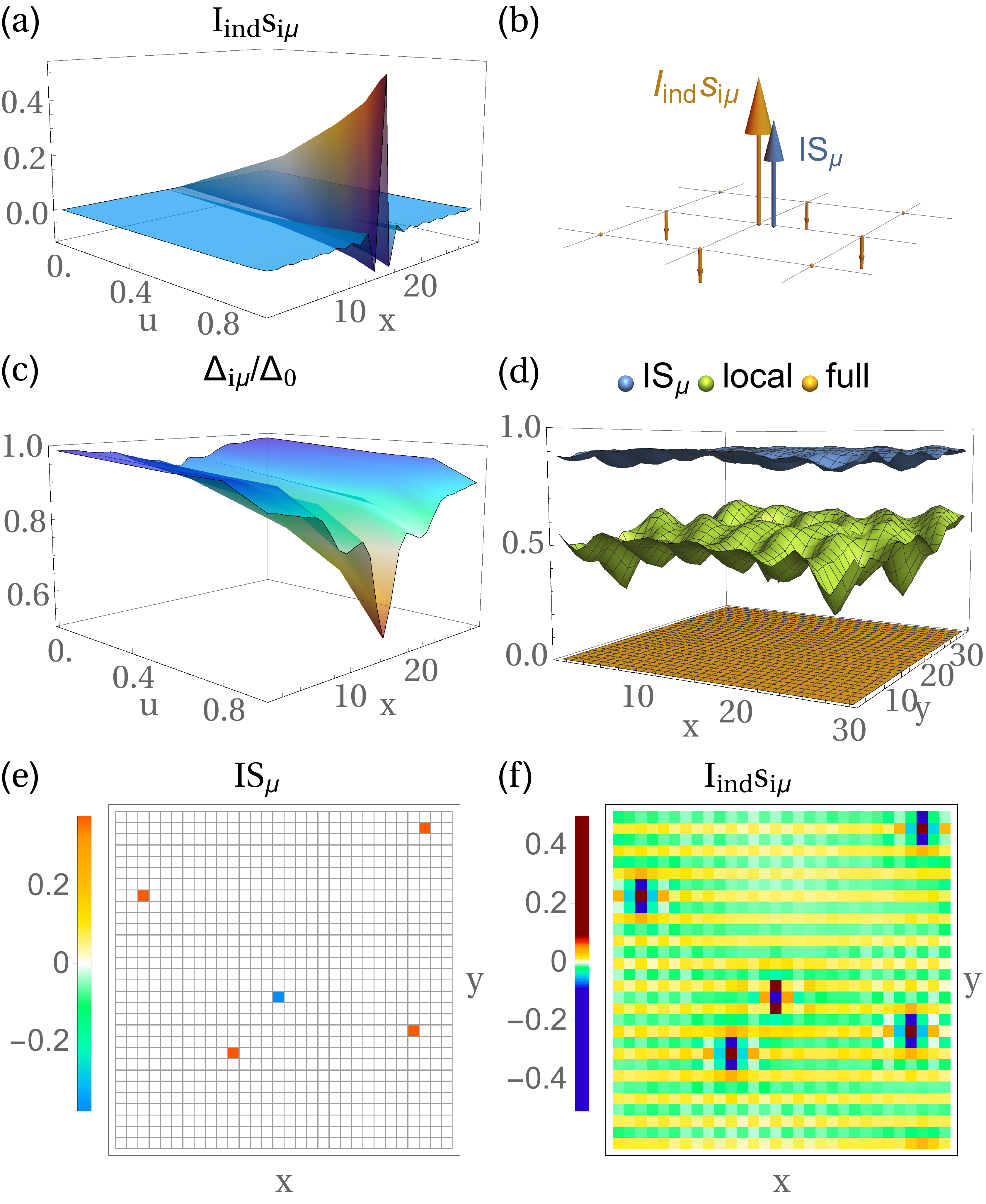}
\end{center}
\caption{{\bf Local and non-local effect of correlations on the impurity response from magnetic disorder.} 
(a) Induced magnetic potential, $I_{ind}s_{\mathbf{i}\mu}$, along a cut through the impurity and as a function of $u$. 
Correlations strongly renormalize the local moments and lead to typical real-space extended magnetic puddles similar to the one shown in (b) where the orange (blue) arrows show the induced (bare) moments for $u=0.97$.  
The resulting disorder potential significantly modifies the suppression of the superconducting order parameter as seen in (c). 
(d) Real-space map of the superconducting order in the presence of $0.5\%$ magnetic disorder. 
The blue surface shows the suppression from only the bare moments, i.e. $u=0$, but self-consistently obtained gaps beyond AG theory. 
Including the local correlation-enhanced magnetic moments leads to the green surface, and only by including both local and non-local effects is superconductivity fully destroyed (orange). 
(e,f) Real-space maps of the (e) bare moments $IS_{\mu}$ and (f) induced magnetic potential $I_{ind}s_{\mathbf{i}\mu}$ for $u=0.97$. 
The correlations strongly enhance the inter-impurity coupling by inducing a LRO magnetic phase in-between the disorder sites, which further competes with superconductivity and efficiently suppresses $T_c$. 
For all results in this figure, the bare moments are the ones used in Fig.~\ref{fig:2} and $\mu=d_{xz}$ orbital. }
\label{fig:3}
\end{figure}

Figure~\ref{fig:2} shows the suppression of $T_c$ as a function of magnetic impurity concentration obtained within our model. 
Without correlations ($u=0$) the $T_c$ suppression follows the curve expected from AG theory. 
The main result of Fig.~\ref{fig:2} is the much faster $T_c$-suppression rate when including Coulomb interactions as seen by comparison of the two curves in Fig.~\ref{fig:2}. 
How may one understand this result which appears at odds with the expectation that correlations screen disorder and limit their damaging effects?\cite{andersen08,garg08,fukushima09,guzman13,vlad15} 
The answer to this question necessitates a deeper understanding of correlation effects at both the {\it local} scale (immediate vicinity of the impurity sites) and {\it non-local} scale (inter-impurity regions). 
Both effects are intimately tied to the fact that magnetic impurity moments induce spin polarizations of the surrounding itinerant electrons $m_{\mathbf{i}\mu}$, which  renormalize the exchange coupling such that $\mathcal{\tilde H}_{imp}={\tilde I}\sum_{\mathbf{i}\mu\sigma}\sigma{\tilde S}_{\mathbf{i}\mu} c_{\mathbf{i}\mu\sigma}^{\dagger}c_{\mathbf{i}\mu\sigma}$, where 
\begin{align}
\label{renormexchange}
 {\tilde I}{\tilde S}_{\mathbf{i}\mu}&= \left[ IS_\mu \delta_{\mathbf{i}\mathbf{i^*}} -\frac{1}{2}\left(U m_{\mathbf{i}\mu}+J\sum_{\nu\neq\mu}m_\mathbf{i\nu} \right) \right] \\\nonumber
 &\equiv \left[ IS_\mu \delta_{\mathbf{i}\mathbf{i^*}} + I_{ind} s_{\mathbf{i}\mu}\right],
\end{align}
is the emergent interaction-generated extended magnetic impurity potential (see SM for more details) generated by the induced part, $ I_{ind} s_{\mathbf{i}\mu}$. 
Focussing first on the {\it local} part of the effective potential, a line-cut of the induced magnetic potential $I_{ind}s_{\mathbf{i}\mu}$ through a single impurity as a function of $u$ is shown in Fig~\ref{fig:3}(a). 
As seen, the extent and amplitude of the resulting magnetic puddle grows significantly with $u$, and results in a real-space structure illustrated in Fig.~\ref{fig:3}(b). 
The renormalized moment is significantly {\it enhanced} at the impurity site, even exceeding the bare moment at large $u$, and exhibits sizable anti-parallel neighboring spins. 
Superconductivity is strongly affected by the additional pair-breaking caused by the enhanced local moments, and therefore the suppression of the order parameter $\Delta_{\mathbf{i}}$ increases accordingly, as shown in Fig.~\ref{fig:3}(c). 
In this way, approaching the magnetic instability has conspicuous local damaging effects on superconductivity. This enhanced local pair-breaking is not, however, the sole reason for the enhanced $T_c$ suppression rate, which also includes a cooperative (non-local) multi-impurity effect.

Indeed, when multiple impurities are included, the correlations among their moments become crucial for lowering the free energy. Specifically, the spin polarized clouds around the impurities prefer to constructively interfere, thereby generating a quasi-long-range ordered magnetic state.~\cite{Mngeese} The inter-impurity regions acquire a resulting finite magnetization due to this enhanced RKKY-like interaction between the impurities. Figures~\ref{fig:3}(e) and \ref{fig:3}(f) compare directly the case in point with $0.5\%$ uncorrelated disorder ($u=0$) versus the correlated situation ($u=0.97)$, respectively. In addition to the local effect discussed above, the system develops $(\pi,0)$ LRO magnetization (see also SM) in agreement with experiments~\cite{carretta16}, constituting the additional non-local competitor to superconductivity. We show in Fig.~\ref{fig:3}(d) a plot of these two separate (local vs. non-local) effects on the superconducting order parameter suppression. The blue surface is the self-consistent solution of $\Delta_{\mathbf{i}}$ of the $u=0$ system shown in Fig.~\ref{fig:3}(e). 
The superconducting order parameter is hardly affected by the bare magnetic potentials, and this is reflected in the correspondingly low $T_c$ suppression of Fig.~\ref{fig:2}. The green surface of Fig.~\ref{fig:3}(d) is the resulting substantially reduced inhomogeneous $\Delta_{\mathbf{i}}$ solution of the gap equation due to the renormalized \emph{local} potentials, cf. Fig.~\ref{fig:3}(b). 
Only when the second non-local magnetic order is also included, superconductivity is completely wiped out as illustrated by the orange surface in Fig.~\ref{fig:3}(d), explaining the physics of the aggressive sub-1 $\%$ $T_c$ suppression rate shown in Fig.~\ref{fig:2}.

Within the above scenario, why does it require an order of magnitude more magnetic disorder to suppress $T_c$ to zero in, for example, Sm-1111 compared to La-1111? We point out two main reasons: 1) OD Sm-1111 exhibits a larger $T_c$ (compared to OD La-1111) (see SM for details), and 2) Consistent with transport studies,~\cite{hess09} Sm-1111 is less correlated than La-1111, and hence described by effective interactions further away from the quantum critical point (QCP) at $u=1$ than La-1111. In the SM we show that indeed 8\% critical amounts of magnetic disorder in OD Sm-1111 is consistent with our modelling. Recently it was shown that Y-substitution for La can similarly shift OD La-1111 away from the QCP and remove the poisoning effect.~\cite{moroni16}

{\it Non-magnetic disorder.} 
We now turn to the discussion of non-magnetic disorder, and again motivate the study by a set of puzzling experimental findings from FeSCs summarized in Figs.~\ref{fig:1}(c,d), which compare the effect on $T_c$ and $T_m$ of Ru ions substituting for Fe in OD La-1111 and Sm-1111\cite{satomi10,sanna_prl,sanna_prb}. 
Ru is isovalent to Fe, and therefore expected to be a source of weak disorder, consistent with the huge amount of $\sim 60\%$ of Ru required to suppress $T_c$, as seen in Fig.~\ref{fig:1}(c). 
An unexpected magnetic phase is induced at intermediate values of Ru content $x$, centered roughly around $x=0.25$, and existing only at a finite span $\Delta x$ of disorder as seen in Fig.~\ref{fig:1}(d). 
The magnetic phase is most pronounced with largest $\Delta x$ and highest $T_m$ in Sm-1111 and only marginally present in La-1111 even though the latter system displays the poisoning effect and hypothetized to be more correlated (than Sm-1111) in the above discussion. Finally we point out the remarkable counterintuitive levelling-off of the $T_c$ suppression rate concomitant with the value of Ru content $x_c$ where magnetic order sets in, as seen most clearly in the case of Sm-1111 in Fig.~\ref{fig:1}(d). 

In order to  capture correctly the effects of large (compositional changing)  amounts of Ru substitution, it is imperative to include the effect of Ru on the bandstructure itself. 
Our first-principles calculations show that the bandwidth roughly doubles with Ru content going from $x=0$ to $x=1$ in both LaFe$_{1-x}$Ru$_x$AsO and SmFe$_{1-x}$Ru$_x$AsO (see SM for details). 
This band-widening effect is accounted for by an overall renormalization of the hopping amplitudes $t_{\mathbf{ij}}^{\mu\nu}\rightarrow(1+x)t_{\mathbf{ij}}^{\mu\nu}$ in Eq.~(\ref{eq:H0}). 
For concreteness, we focus initially on a case with correlations of intermediate strength, $u=0.7$, since this seems relevant for e.g. Sm-1111 which exhibits the most pronounced disorder-induced magnetic phase as shown in Fig.~\ref{fig:1}(d). 
Consistent with first-principles calculations,~\cite{nakamura11} we model the random collection of non-magnetic Ru ions by a set of weak point-like scatterers with $V_{\mu}=0.03$ eV on all orbitals but allow for a phenomenological tuning of the potential on the $d_{3z^2-r^2}$ orbital ($V_{d_{3z^2-r^2}}=0.7$ eV). 
The latter is necessary in order to locally stabilize magnetism for the particular band utilized in this work.~\cite{ikeda10} 

\begin{figure}[]
\begin{center}
\includegraphics[width=0.9\columnwidth]{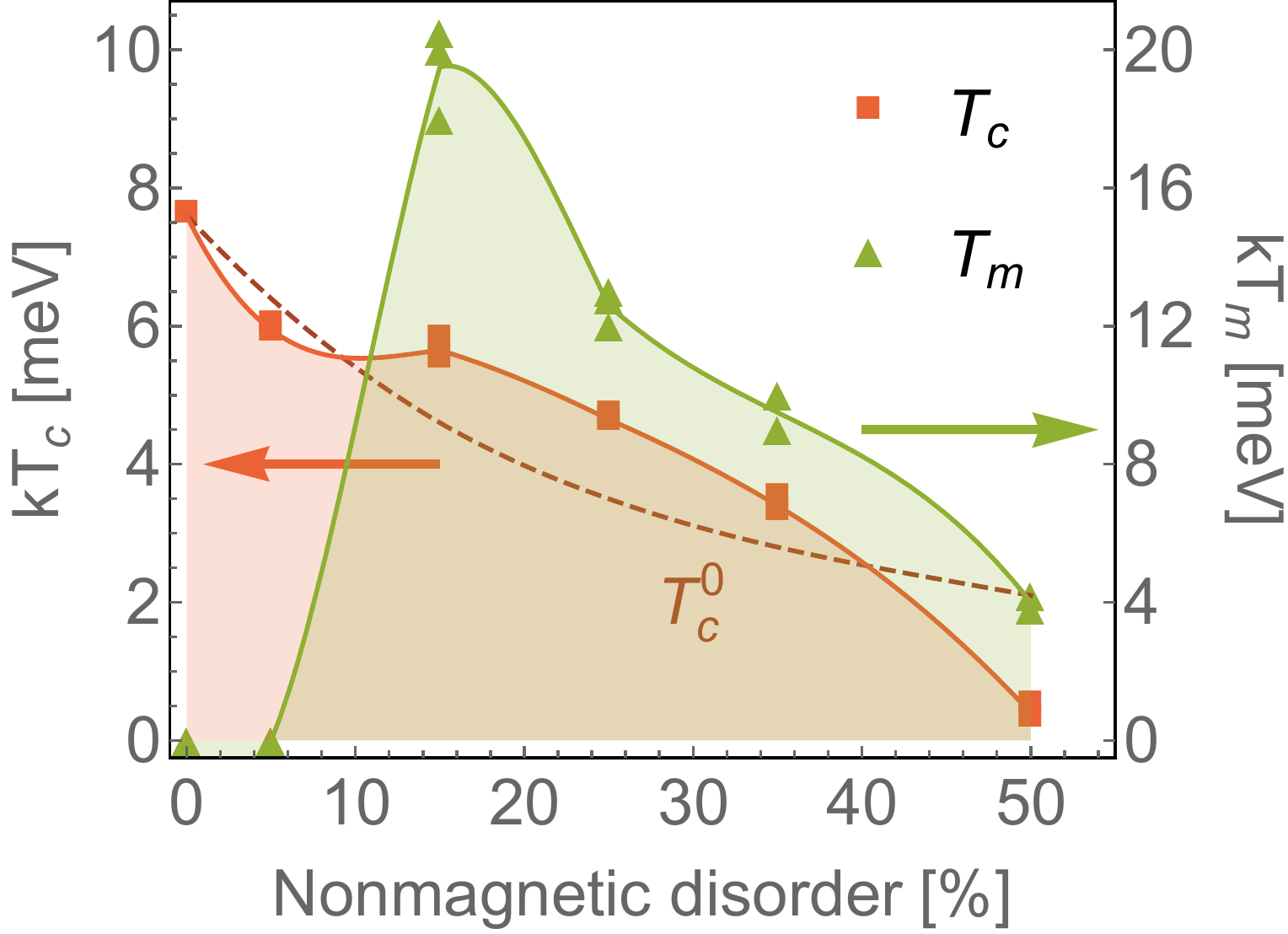}
\end{center}
\caption{{\bf Critical temperatures $T_c$ and $T_m$ as a function of non-magnetic impurity concentration $x$.} 
The critical temperature $T_c$ versus disorder concentration (red squares). 
The dashed curve shows the $T_c$ for the clean system $T_c^0$ where only the band-widening effect has been included (i.e. no disorder), effectively reducing the pairing strength by $\tilde \Gamma \equiv \Gamma/(1+x)$. As seen there is a region $\Delta x$ of disorder concentration ($10\% \lesssim x \lesssim 40\%)$ where the bound state effect (see main text) has enhanced $T_c$ for the disordered case as compared to the clean system.  This regime is characterized by the existence of a bulk magnetic phase (green triangles) induced by the disorder and is seeded by favorable local impurity structures as explained in Fig.~\ref{fig:5}. Note the kink in $dT_c/dx$ and a {\it reduced} $T_c$ suppression rate around $x\sim 10\%$ concomitant with the onset of volume-full magnetic order.
}\label{fig:4}
\end{figure}

\begin{figure}[]
\begin{center}
\includegraphics[width=0.99\columnwidth]{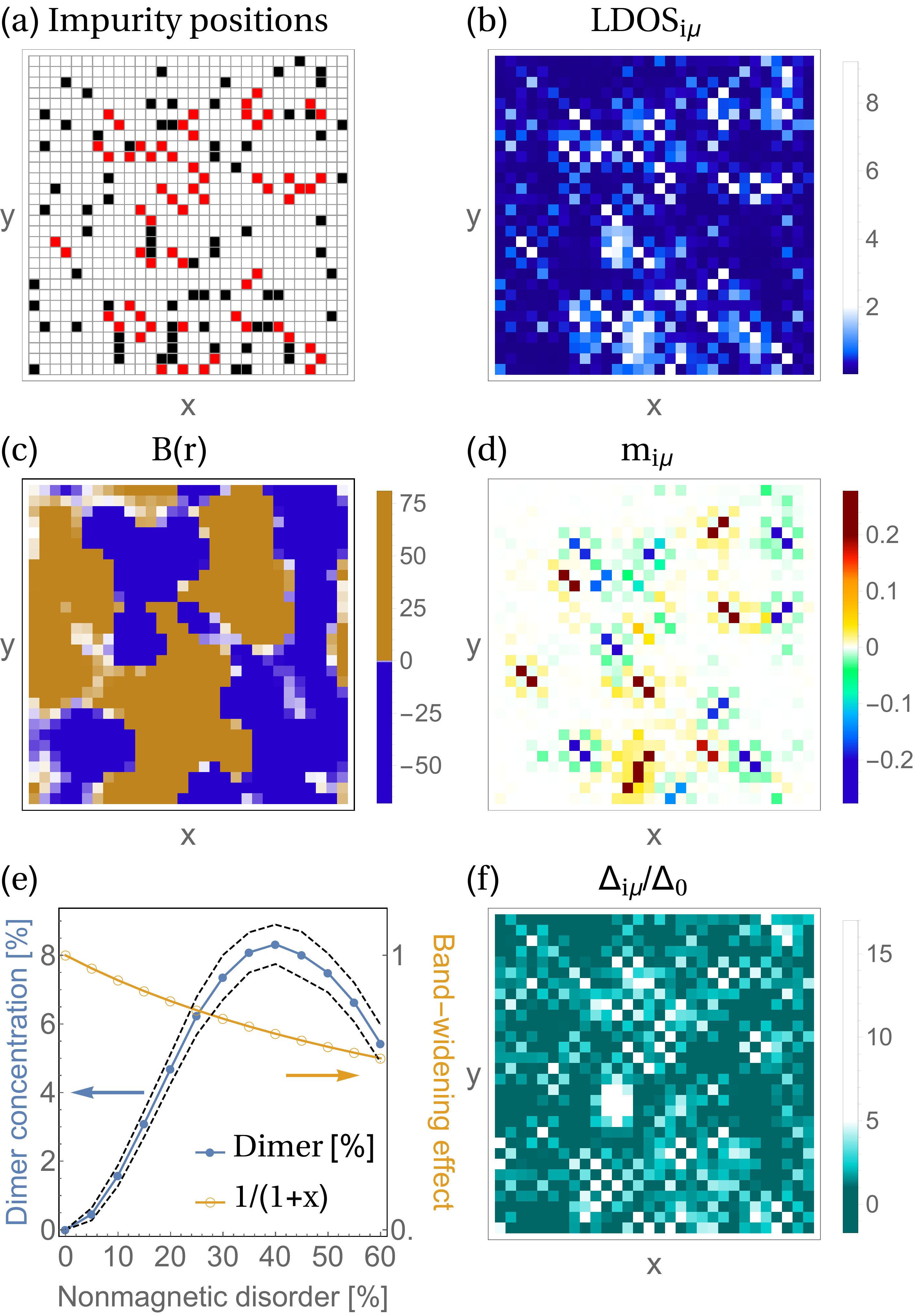}
\end{center}
\caption{{\bf Effect of impurity dimers on $T_m$ and $T_c$.} 
(a) Black and red tiles both indicate the positions of a random set of $15\%$ disorder. 
The red tiles highlight favorable dimer-like arrangements, defined by all the impurity sites with an occupied NNN site but not more than one occupied NN site. 
(b) Real-space map of the LDOS of the $d_{3z^2-r^2}$ orbital at $T>T_c$ at the Fermi level, revealing explicitly the correlation between the brightest sites (largest LDOS) with the red dimer sites in (a). 
(c) Local dipolar field $B(r)=\sum_{\mathbf i} \frac{m_{\mathbf i}}{|\mathbf{r}_{\mathbf i}|^3}$ ($\mathbf{r}_{\mathbf i}$ is the distance between the muon site $\mathbf{r}$ and the moment position $m_{\mathbf i}$ of the itinerant electrons) felt by muons with orange (blue) color indicating regions with field strength larger (smaller) than $0.5$mT ($-0.5$mT). (d) Real-space map of the dimer-induced magnetization of the $d_{3z^2-r^2}$ orbital which dominates the total magnetization. From a comparison to (b) it is evident that the LDOS enhancement near the dimers freeze magnetic order in their vicinity. 
(e) Dimer density (blue dots) and the bandwidth renormalization parameter $1/(1+x)$ (orange circles) as a function of disorder concentration $x$. 
(f) Superconducting order parameter of the $d_{3z^2-r^2}$ orbital $\Delta_{{\mathbf i} d_{3z^2-r^2}}/\Delta_{d_{3z^2-r^2}}^0$ relative to its value in the clean system, revealing remarkable order-of-magnitude local enhancements in the vicinity of the impurity dimers.
}
\label{fig:5}
\end{figure}

Figure~\ref{fig:4} shows the resulting critical temperatures $T_c$ and $T_m$ as a function of $x$. 
As seen, in addition to a much slower $T_c$ suppression rate as compared to Fig.~\ref{fig:2}, a magnetic phase centered around $x\sim 25 \%$ is generated above a certain concentration $x_c$ of Ru ions. 
As a function of $x$, $T_c$ exhibits an initial drop, but, interestingly, the induction of the magnetic phase does not enhance the $T_c$ suppression rate as expected from naive competitive considerations, but rather seems to {\it further stabilize} superconductivity. 
The origin for these unconventional disorder effects is explained in Fig.~\ref{fig:5} and the associated caption. 
In essence, the emergence of favorable impurity clusters (dimers and trimers, Fig.~\ref{fig:5}(a)) lead to substantial LDOS enhancements of the $d_{3z^2-r^2}$ orbital (Fig.~\ref{fig:5}(b)), which drive both 1) induced magnetization (Fig.~\ref{fig:5}(d)) through local crossings of the Stoner instability,~\cite{andersen1,gastiasoro13} and 2) an associated enhancement of the superconducting order parameter $\Delta_{\mathbf{i}d_{3z^2-r^2}}$ (Fig.~\ref{fig:5}(f)). 
Through inter-orbital couplings the boost of $\Delta_{\mathbf{i} d_{3z^2-r^2}}$ near the dimers is enough to cause the support for the entire superconducting condensate evident in Fig.~\ref{fig:4} at intermediate disorder content $\Delta x$, where the enhanced pairing overcompensates the pair-breaking effect of both the disorder and the induced magnetic phase.

The dimer-induced LDOS enhancement mechanism naturally explains the increase of $T_m$ starting at intermediate values of impurity concentration $x_c\sim 10 \%$, since no favorable impurity clusters are present below $x_c$. 
As the concentration of disorder increases, more dimer-like structures with high LDOS form, and the system eventually acquires a large enough magnetic volume fraction to support a non-zero $T_m$. 
Specifically, $T_m$ is defined identically to the experimental $\mu$SR definition by the highest $T$ exhibiting a $50\%$ magnetic volume fraction. 
A site is defined to contribute to the volume fraction if its internal dipolar local field exceeds $|0.5|$mT~\cite{sanna_prl,sanna_prb}. 
In the case of $15 \%$ disorder discussed in Fig.~\ref{fig:5} we find a nearly saturated volume fraction as shown in panel \ref{fig:5}(c), in agreement with experiments~\cite{sanna_prl,sanna_prb}. 
For more details on the definition of $T_m$ and the resulting short-range magnetic structure induced by the dimers, we refer to the SM. 
From Fig.~\ref{fig:5}(e), showing the dimer concentration as a function of $x$ one expects a max $T_m$ near $x\sim40\%$, however, the band-widening effect $W\rightarrow(1+x)W$ results in lowered effective Coulomb correlations, and pushes the magnetic dome to lower $x$. 
Thus, the position of the induced magnetic dome is a compromise between the dimer-enhanced LDOS and the weakening of correlations due to band-widening. 
The resulting $x$-dependence of both $T_c$ and $T_m$ seen in Fig.~\ref{fig:4} appear in excellent overall agreement with the experimental results shown in Fig.~\ref{fig:1}(d). 

Returning to the discussion of the distinction between the two 1111 materials shown in Fig.~\ref{fig:1}(c,d), a remaining question is the origin of the significantly smaller induced magnetic phase in La-1111 compared to Sm-1111. La-1111 exhibits the poisoning effect explained above by a larger $u$ (compared to Sm-1111) in this material, and accordingly one may naively expect the induced magnetic phase to be even more pronounced for La-1111 than for Sm-1111. 
However, the larger correlations also act to screen the non-magnetic disorder which results in lower effective potentials (see SM for details) on the $e_g$ orbitals which then become unable to cause the LDOS enhancements shown in Fig.~\ref{fig:5}. 
The modified potentials simply shift the bound-state structure away from the Fermi level, locally weakening the Stoner condition and thereby overcompensating the larger $u$ as explained in the SM.

We end by pointing out that the main effects discussed in this work, i.e. the poisoning effect by magnetic disorder and the resilience of superconductivity to nonmagnetic disorder and its induced magnetization, are not a pecularity of iron-based systems, but rather quite general effects expected to exist in multi-orbital correlated superconducting systems. 
By tuning other materials close to a magnetic instability, for example, magnetic disorder should exhibit a similar aggressive $T_c$-suppression rate. Likewise, when nonmagnetic disorder leads to large enough LDOS enhancements of orbitals that do not dominate the spectral weight in the clean system near the Fermi level, a disorder-induced coexistence phase of magnetism and superconductivity is expected to occur. 
Our findings also serve as a warning to draw strong conclusions about the pairing symmetry based on $T_c$-suppression rates of unconventional correlated systems without detailed theoretical modelling beyond standard AG-theory.

\section{Acknowledgements}

We thank P. Carretta, M. H. Christensen, F. Hammerath, P. J. Hirschfeld, D. Inosov, A. Kreisel, M. Moroni, R. di Renzi, S. Sanna, and D. J. Scalapino for useful discussions. M.N.G. and B.M.A, acknowledge support from Lundbeckfond fellowship (grant A9318). F. B. acknowledges partial support from PRIN Grant No. 2012X3YFZ2-004 and from FP7-EU project SUPER-IRON (No. 283204).


\pagebreak
\widetext
\begin{center}
\textbf{\large Supplemental Materials: "Unconventional disorder effects in correlated superconductors"}
\end{center}
\setcounter{equation}{0}
\setcounter{figure}{0}
\setcounter{table}{0}
\setcounter{page}{1}
\makeatletter
\renewcommand{\theequation}{S\arabic{equation}}
\renewcommand{\thefigure}{S\arabic{figure}}
\renewcommand{\bibnumfmt}[1]{[S#1]}
\renewcommand{\citenumfont}[1]{S#1}

Here, we provide computational details of the results presented in the main part of the paper. We elaborate both on the applied band structure, and the mean-field decoupled Hamiltonian in real space. We outline the details of our Abrikosov-Gor'kov calculations in orbital space. Finally, we discuss the role of magnetic (non-magnetic) disorder in Sm-1111 (La-1111), and show how the relevant experimental findings for those cases may also be naturally reconciled within the correlated disorder scenario presented in the main part of the paper.

\section{Model}

The starting Hamiltonian defined on a two-dimensional lattice is given by
\begin{equation}
 \mathcal{H}=\mathcal{H}_{0}+\mathcal{H}_{int}+\mathcal{H}_{BCS}+\mathcal{H}_{imp},
\end{equation}
describes a superconducting system in the presence of correlations and disorder.

\begin{figure}[b]
\begin{center}
\includegraphics[width=0.4\columnwidth]{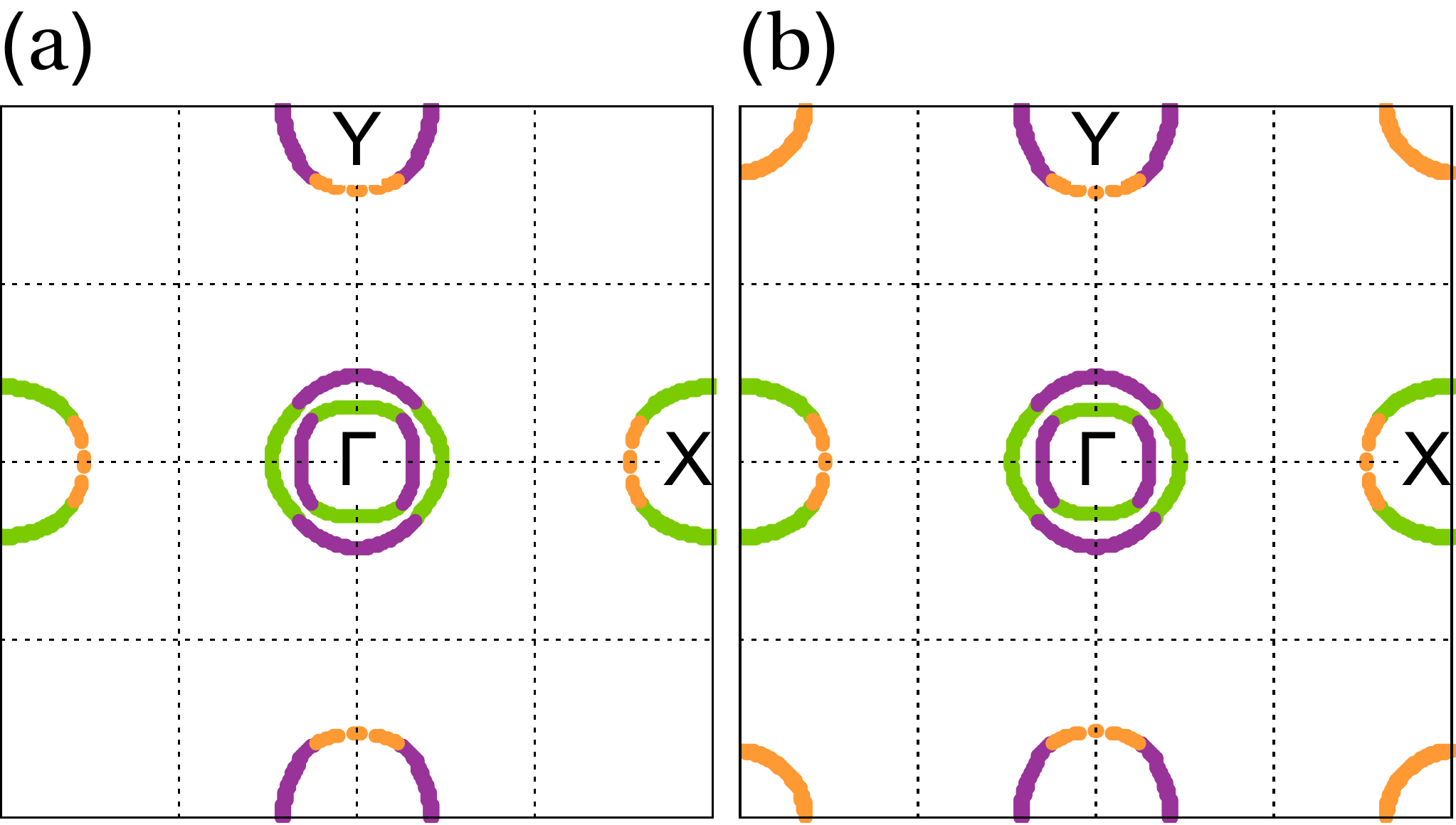}
\end{center}
\caption{Fermi surface used to model (a) La-1111 and (b) Sm-1111. Different colors represent the main orbital character of the bands (purple: $d_{xz}$; green: $d_{yz}$; orange: $d_{xy}$).  
As the pnictogen height decreases (Sm $\rightarrow$ La) the $d_{xy}$ $\gamma$ pocket at $(\pi,\pi)$ disappears from the Fermi surface. 
In our modelling, the presence of this pocket is controlled by the NN $d_{xy}$ hopping parameter, $ t_{xy}^{La}=1.25t_{xy}^{Sm}$.}
\label{fig:sm1}
\end{figure}

We use a five-orbital tight-binding band relevant to the 1111 pnictides~\cite{Sikeda10} 
\begin{equation}
\mathcal{H}_{0}=\sum_{\mathbf{ij},\mu\nu,\sigma}t_{\mathbf{ij}}^{\mu\nu}c_{\mathbf{i}\mu\sigma}^{\dagger}c_{\mathbf{j}\nu\sigma}-\mu_0\sum_{\mathbf{i}\mu\sigma}n_{\mathbf{i}\mu\sigma}.
\end{equation}
We stress that for the 1111 systems a two-dimensional model should be appropriate since the dispersion along the $k_z$ direction is essentially absent.\cite{Sikeda10} 
Figure~\ref{fig:sm1} shows the Fermi surface and its main orbital character.
The presence of the $\gamma$ pocket at $(\pi,\pi)$ was found to depend on the pnictogen height~\cite{Skuroki09}, which can be controlled by the nearest-neighbor (NN) hopping parameter $t_{xy}$. 
Here, we use this result to model the difference between La-1111 (without the $\gamma$ pocket) and Sm-1111 (with the $\gamma$ pocket) by setting $ t_{xy}^{La}=1.25t_{xy}^{Sm}$. 
The resulting Fermi surfaces are shown in Fig.~\ref{fig:sm1}. 

After a mean-field decoupling, the interacting multi-orbital Hubbard interaction becomes
\begin{equation}
\label{eq:hintmf}
 \mathcal H_{int}^{MF}=\sum_{\mathbf{i},\mu\neq\nu,\sigma}[U n_{\mathbf{i}\mu\overline{\sigma}}+U'n_{\mathbf{i}\nu\overline{\sigma}}+
(U'-J) n_{\mathbf{i}\nu\sigma}]\hat{c}_{\mathbf{i}\mu\sigma}^{\dagger}\hat{c}_{\mathbf{i}\mu\sigma},
\end{equation}
where $n_{\mathbf{i}\mu\sigma}\equiv \langle\hat{c}_{\mathbf{i}\mu\sigma}^{\dagger}\hat{c}_{\mathbf{i}\mu\sigma}\rangle$.
We apply the spin and orbital rotational invariance relations $U'=U-2J$ and $J'=J$ throughout this work, and additionally set $J=U/4$. 

Superconductivity is included by a BCS-like term
\begin{equation}
 \mathcal{H}_{BCS}=-\sum_{\mathbf{i}\neq \mathbf{j},\mu\nu}[\Delta_{\mathbf{ij}}^{\mu\nu}c_{\mathbf{i}\mu\uparrow}^{\dagger}c_{\mathbf{j}\nu\downarrow}^{\dagger}+\mbox{H.c.}],
\end{equation}
with $\Delta_{\mathbf{ij}}^{\mu\nu}=\sum_{\alpha\beta}\Gamma_{\mu\alpha}^{\beta\nu}(\mathbf{r_{ij}})\langle\hat{c}_{\mathbf{j}\beta\downarrow}\hat{c}_{\mathbf{i}\alpha\uparrow}\rangle$ being the superconducting order parameter, and $\Gamma_{\mu\alpha}^{\beta\nu}(\mathbf{r_{ij}})$ denoting the effective pairing strength between sites (orbitals) $\mathbf{i}$ and $\mathbf{j}$ ($\mu$, $\nu$, $\alpha$ and $\beta$). 
In agreement with a general $s^\pm$ pairing state, we include next-nearest neighbor (NNN) intra-orbital pairing, $\Gamma_{\mu}\equiv\Gamma_{\mu\mu}^{\mu\mu}(\mathbf{r_{nnn}})$. 
Additionally, the standardly obtained reduced pairing vertex of the $e_g$ orbitals (see for example Ref.~\onlinecite{GastiasoroLiFeAs2013}) is accounted for by reducing them by roughly a factor of two: $\Gamma_{t_{2g}}=0.293$ eV and $\Gamma_{e_g}=0.5\Gamma_{t_{2g}}$.
This reduction limits the $T_c$ enhancement found in the nonmagnetic disorder case (see Fig.4 in main text), where the $e_g$ orbitals play an important role. 
By contrast, in the case of magnetic disorder the reduction in the $e_g$ orbital pairing does not influence the results since both superconductivity and induced long-range polarization are mainly determined by the $t_{2g}$ orbitals.

The last term in the Hamiltonian introduces disorder in the system.
Non-magnetic and magnetic disorder give rise to terms of the form $\mathcal{H}_{imp}=\sum_{\mu\{\mathbf{i^*}\}}V_{\mu} n_{\mathbf{i^*}\mu}$ and $\mathcal{H}_{imp}=I\sum_{\{\mathbf{i^*}\}\mu\sigma}\sigma S_\mu c_{\mathbf{i^*}\mu\sigma}^{\dagger}c_{\mathbf{i^*}\mu\sigma}$, respectively. 
Here $V_{\mu}$ ($S_\mu$) denotes the impurity potential (magnetic moment) in orbital $\mu$ at the disorder sites given by the set $\{\mathbf{i^*}\}$ coupled to the charge (spin) density of the itinerant electrons.

By using the spin-generalized Bogoliubov transformation,
\begin{align}
\label{eq:bogtrans}
\hat{c}_{\mathbf{i}\mu\sigma}&=\sum_{n}(u_{\mathbf{i}\mu\sigma}^{n}\hat{\gamma}_{n\sigma}+v_{\mathbf{i}\mu\sigma}^{n*}\hat{\gamma}_{n\overline{\sigma}}^{\dagger}),
\end{align}
we arrive to the Bogoliubov-de Gennes (BdG) equations 
\begin{eqnarray}
\begin{pmatrix}
\hat{\xi}_{\uparrow} & \hat{\Delta}_{\mathbf{ij}}\\
\hat{\Delta}_{\mathbf{ji}}^{*} & -\hat{\xi}_{\downarrow}^{*} 
\end{pmatrix}
\begin{pmatrix}
 u^{n} \\ v^{n} 
\end{pmatrix}=E_{n}
\begin{pmatrix}
 u^{n} \\ v^{n} 
\end{pmatrix}.
\end{eqnarray}
The transformation
$\begin{pmatrix}
   u_{\uparrow}^{n} && v_{\downarrow}^{n} && E_{n\uparrow}
 \end{pmatrix}
\rightarrow
\begin{pmatrix}
 v_{\uparrow}^{n*} && u_{\downarrow}^{n*}&&-E_{n\downarrow}
\end{pmatrix}$
maps two of the equations onto the other two and thus we drop the spin index from the eigenvectors and eigenstates of the BdG equations.
The matrix operators are defined as:
\begin{align}
 \hat{\xi}_{\sigma}u_{\mathbf{i}\mu}&=\sum_{j\nu}t_{\mathbf{ij}}^{\mu\nu}u_{\mathbf{j}\nu}+\sum_{\mu\neq\nu}\left[-\mu_0+\Omega\delta_{\mathbf{ii^*}}\delta_{\mu\nu}+U n_{\mathbf{i}\mu\overline{\sigma}}+U'n_{\mathbf{i}\nu\overline{\sigma}}+(U'-J)n_{\mathbf{i}\nu\sigma} \right] u_{\mathbf{i}\mu},\\\nonumber
\hat{\Delta}_{\mathbf{ij}}^{\mu\nu}u_{\mathbf{i}\mu}&=-\sum_{\mathbf{j}\nu}\Delta_{\mathbf{ij}}^{\mu\nu}u_{\mathbf{j}\nu},
\end{align}
where $\Omega=V_{\mu}$ for nonmagnetic disorder and $\Omega=\sigma I S_{\mu}$ for magnetic disorder.
The five-orbital BdG equations are solved on $30\times30$ lattices with stable solutions found through iterations of the following self-consistency equations 
\begin{align}
\label{eq:bdg}
  n_{\mathbf{i}\mu\uparrow} &=\sum_{n}|u_{\mathbf{i}\mu}^{n}|^{2}f(E_{n}),\\\nonumber
 n_{\mathbf{i}\mu\downarrow} &=\sum_{n}|v_{\mathbf{i}\mu}^{n}|^{2}(1\!-\!f(E_{n})),\\\nonumber
\Delta_{\mathbf{ij}}^{\mu}&=\Gamma_{\mu}\sum_{n}u_{\mathbf{i}\mu}^{n}v_{\mathbf{j}\nu}^{n*}f(E_{n}),
\end{align} 
where $\sum_n$ denotes summation over all eigenstates $n$. We stress that the solutions are fully unrestricted and allowed to vary on all lattice sites and orbitals.
The superconducting order parameter shown in the main manuscript is the bond averaged singlet component:
\begin{equation}
 \Delta_{\mathbf i\mu}=\frac{1}{4}\sum_{\mathbf j}\frac{1}{2}(\Delta_{\mathbf{ij}}^{\mu}+\Delta_{\mathbf{ji}}^{\mu})
\end{equation}
where $\mathbf j$ are four nearest neighbors.
The inclusion of several impurities leads to a spatially varying order parameter $\Delta_{\mathbf{i}\mu}$, and lowers the transition temperature $T_c$ at which a non-zero solution of the gap equation exists. 
Eventually, for a sufficiently high critical concentration of impurities superconductivity is destroyed at all sites, i.e. $\Delta_{\mathbf{i}\mu}=0$ , and $T_c=0$.

\section{Interaction-generated nonmagnetic and magnetic potentials around disorder}
\label{sec:induced_coup}

\begin{figure}
\begin{center}
\includegraphics[width=0.99\columnwidth]{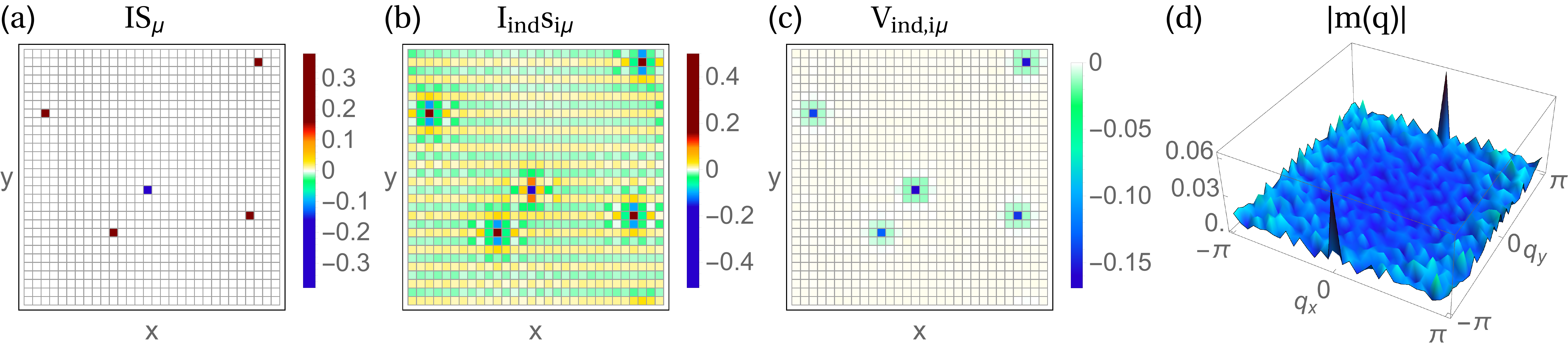}
\end{center}
\caption{(a) Bare magnetic impurities $IS_{\mu}=0.38$ eV and the correlation-induced (b) magnetic potential $I_{ind} s_{\mathbf{i}\mu}$ and (c) nonmagnetic potential $V_{ind,\mathbf{i}\mu}$ for the $d_{xz}$ orbital. 
(d) Fourier transform of the total magnetization. For all panels in this figure $u=0.97$.}
\label{fig:isinduced}
\end{figure}

In order to shed light on the interaction-generated extended potentials we rewrite Eq.~\eqref{eq:hintmf} in terms of the charge density $n_{\mathbf{i}\mu}$ and spin density $m_{\mathbf{i}\mu}$ fields, by using $n_{\mathbf{i}\mu\sigma}=(n_{\mathbf{i}\mu}+\sigma m_{\mathbf{i}\mu})/2$  
\begin{equation}
\label{eq:hintmf2}
 \mathcal H_{int}^{MF}=\frac{1}{2}\sum_{\mathbf{i}\mu\neq\nu\sigma}\left[ U n_{\mathbf{i}\mu}+(2U-5J)n_{\mathbf{i}\nu} \right]\hat{c}_{\mathbf{i}\mu\sigma}^{\dagger}\hat{c}_{\mathbf{i}\mu\sigma}
 -\frac{1}{2}\sum_{\mathbf{i}\mu\neq\nu\sigma}\sigma\left[U m_{\mathbf{i}\mu}+J m_{\mathbf{i}\nu}\right]\hat{c}_{\mathbf{i}\mu\sigma}^{\dagger}\hat{c}_{\mathbf{i}\mu\sigma}.
\end{equation}
The effect of impurities on both charge and spin densities is given by $n_{\mathbf{i}\mu}=n_{\mathbf{i}\mu}^0+\Delta n_{\mathbf{i}\mu}$ and $m_{\mathbf{i}\mu}=m_{\mathbf{i}\mu}^0+\Delta m_{\mathbf{i}\mu}$, where $n_{\mathbf{i}\mu}^0$ and $m_{\mathbf{i}\mu}^0$ are the fields of the disorder-free system ($m_{\mathbf{i}\mu}^0=0$ throughout this study) and $\Delta n_{\mathbf{i}\mu}$ and $\Delta m_{\mathbf{i}\mu}$ the disorder-induced changes.  
Introducing these expressions in Eq.~\eqref{eq:hintmf2} we obtain extended impurity-like terms $V_{ind,\mathbf{i}\mu}$ and $I_{ind} s_{\mathbf{i}\mu}$ that result in the following effective disorder potentials   
 \begin{align}
  {\tilde V}_{\mathbf{i}\mu}&= \left[ V_\mu \delta_{\mathbf{i}\mathbf{i^*}} +\frac{1}{2}\left(U \Delta n_{\mathbf{i}\mu}+(2U-5J)\sum_{\nu\neq\mu}\Delta n_{\mathbf{i}\nu}\right) \right] \equiv \left[ V_\mu \delta_{\mathbf{i}\mathbf{i^*}} + V_{ind,\mathbf{i}\mu}\right],\\\nonumber
 {\tilde I}{\tilde S}_{\mathbf{i}\mu}&= \left[ IS_\mu \delta_{\mathbf{i}\mathbf{i^*}} -\frac{1}{2}\left(U \Delta m_{\mathbf{i}\mu}+J\sum_{\nu\neq\mu}\Delta m_{\mathbf{i}\nu} \right) \right] \equiv \left[ IS_\mu \delta_{\mathbf{i}\mathbf{i^*}} + I_{ind} s_{\mathbf{i}\mu}\right],
 \end{align}
where $V_\mu$ and $IS_\mu$ are the bare nonmagnetic and magnetic potentials, respectively. 
Thus, in the presence of interactions ($u\neq 0$), the field modulations $\Delta n_{\mathbf{i}\mu}$ and $\Delta m_{\mathbf{i}\mu}$ induced by disorder give rise to effective nonmagnetic and magnetic potentials, respectively.
An example relevant for bare magnetic disorder is shown in Fig.~\ref{fig:isinduced}. As seen, besides the induced magnetic potential shown in Fig.~\ref{fig:isinduced}(b), a concomitant induced nonmagnetic potential is generated from charge density modulations, shown in Fig.~\ref{fig:isinduced}(c). For the superconducting state studied in the main part of the paper, the pair-breaking effect of the charge potential is weak and thus most of the correlation effects arise from the induced magnetic local and non-local components. A Fourier transform of the total magnetization is displayed in Fig.~\ref{fig:isinduced}(d), clearly showing the dominant $(0,\pi)$ LRO of the induced magnetization arising from the regions in-between the impurity sites.

We point out that early local paramagnon theories relevant e.g. to disorder in metallic Pd have also discussed effects of Stoner enhanced susceptibilities.\cite{lederer68} 

\section{Bandwidth increase through Ru substitution}

\begin{figure}[t]
 \begin{center}
  \includegraphics[width=0.45\textwidth]{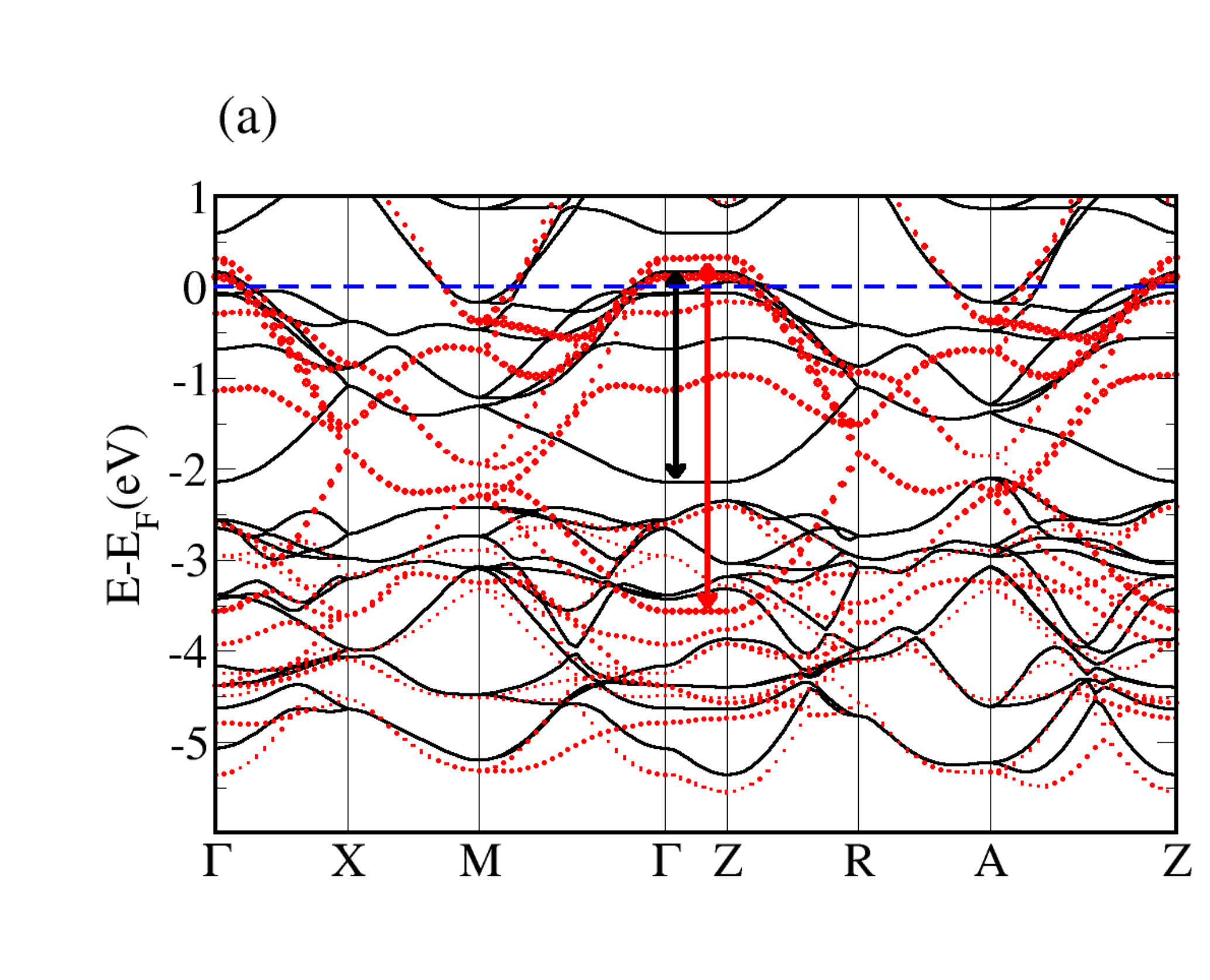}
  \includegraphics[width=0.45\textwidth]{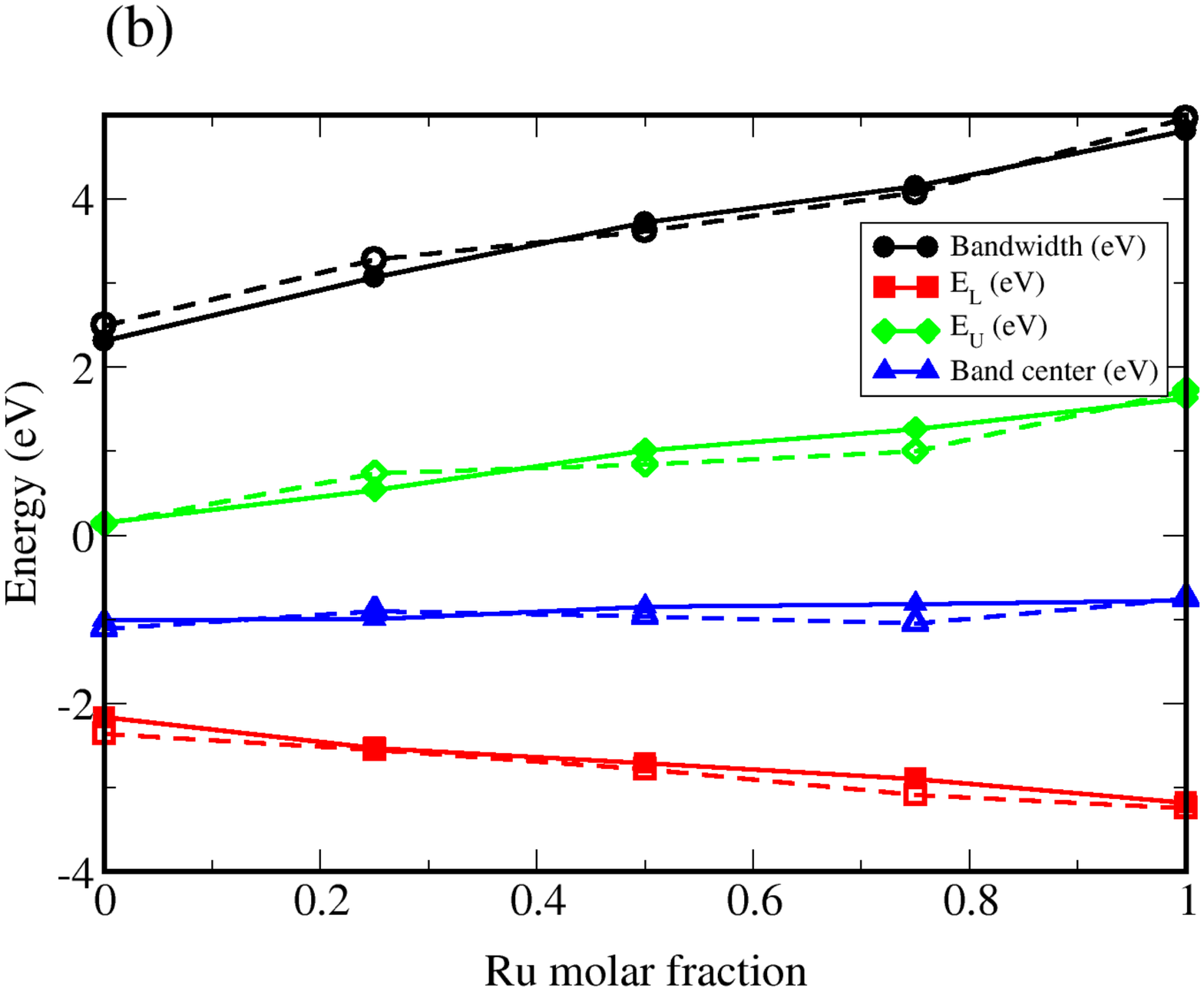}
 \end{center}
    \caption{
    (a) Band structures for LaFeAsO (lines) and LaRuAsO (dots). 
    The size of the dots is proportional to the weight of the Ru-4$d$ orbitals.
    The LaRuAsO band structure is rescaled to fit the first Brillouin zone of LaFeAsO (see text).
    (b) Band parameters for transition metal-related $d$-bands. 
    Solid lines and filled symbols refer to LaFe$_{1-x}$Ru$_x$AsO, while dashed lines and open symbols refer to SmFe$_{1-x}$Ru$_x$AsO.} 
    \label{fig:sm2}
\end{figure}

In Fig.~\ref{fig:sm2}(a), we compare the band structures of LaFeAsO and LaRuAsO to understand the effect of Ru substitution.
The first-principles calculations of the electronic structure were performed within the density functional theory using the full-potential linearized augmented plane-wave  method with the addition of local-orbital basis functions as implemented in the WIEN2K code.~\cite{LocalOrbitals,LocalOrbitals2,wien2k} For the exchange and correlation functional we use the generalized gradient approximation
(GGA) of Perdew, Burke, and Ernzerhof in its revised form.~\cite{PBEsol} We used muffin-tin radii of 2.40 $a_0$ for Sm and La, 2.20 $a_0$ for Fe and Ru, 2.0 $a_0$ for As, and 1.90 $a_0$ for O.

To help the comparison of the results, we rescaled the abscissas in Fig.~\ref{fig:sm2}(a) to fit the band structure of  LaRuAsO with the Brillouin zone for the LaFeAsO system.
The band structure of LaFeAsO (solid lines) is characterized by a valence band originating from  Fe-3$d$ orbitals.
We see that the band ranges from 0.15 to -2.15 eV and is separated from the As-4$p$ band by a small pseudo gap. 
The width of the Fe-3$d$ band is  2.3 eV. 
The short (black) arrow in Fig.~\ref{fig:sm2}(a) shows the estimated width of the band from its topmost $d_{xz}/d_{yz}$ state to the lowermost $d_{x^2-y^2}$ dominated band.
The LaRuAsO and LaFeAsO band structures clearly differ in the dispersion of the Ru-4$d$ orbitals related band. 
To help the readability of the band structure in Fig.~\ref{fig:sm2}(a), we used the so-called fat-bands representation, where the size of the dots is proportional to the weight of the Ru-4$d$ orbitals. 
We see that Ru-4$d$ states span over a range of $\sim$ 4 eV.
In Fig.~\ref{fig:sm2}(a), it is still possible to identify the states with the $d_{xz}/d_{yz}$ and $d_{x^2-y^2}$ characters at $\Gamma$ to define the width of the valence band in LaRuFeAs. 
The long (red) arrow shows the estimated width of the Ru-4$d$ band. 

In order to provide a quantitative estimate of the band width as a function of Ru content $x$ (in e.g. LaFe$_{1-x}$Ru$_x$AsO), we introduce a criterion based on the density of states projected onto the Fe and Ru $d$-orbitals to define the $d$ band width. 
We define the function $P(E)$ by
\begin{equation}
P(E)=\int_{-\infty}^{E} [ (1-x)D_{\rm Fe}(\varepsilon)+xD_{\rm Ru}(\varepsilon)] d\varepsilon,
\end{equation}  
where $D_{\rm Fe}(\varepsilon)$ and  $D_{\rm Ru}(\varepsilon)$ are the values of the projected density of states (PDOS) onto the Fe-3$d$ and Ru-4$d$ orbitals, respectively.
The physical meaning of $P(E)$ is the amount of electron density for energies below $E$ coming from $d$ states of Fe and Ru. 
We use $P(E)$ to define unambiguously the band width in {\it R}Fe$_{1-x}$Ru$_x$AsO. 
We find that $P(E)$ in LaFeAsO is 1.515 for $E=-2.15$ eV and 6.3 for $E=0.15$ eV.
We define the minimum of the transition metal-related $d$-bands, $E_{\rm L}$, as the value that fulfills the relation $P(E_{\rm L})=1.515$, and the maximum, $E_{\rm U}$, as the value for which $P(E_{\rm U})=6.3$.
With this criterion, we computed the lower and upper limits of the $d$ related band in {\it R}Fe$_{1-x}$Ru$_x$AsO.  

In Fig.~\ref{fig:sm2}(b), we show the energies of the upper and lower edges of the transition metal-related band; the band center defined as the average $E_{\rm av}=\frac{1}{2}(E_{\rm U}+E_{\rm L})$; the bandwidth $E_W=E_{\rm U}-E_{\rm L}$.
Fig.~\ref{fig:sm2}(b) shows that the La vs. Sm substitution does not influence the position and the width of the transition metal $d$ band. 
The band center energy $E_{\rm av}$  is weakly dependent of Ru concentration. 
The bandwidth increases from 2.3 to 4.8 with Ru content. 
This is the most relevant effect of Ru substitution on the band structure of {\it R}Fe$_{1-x}$Ru$_x$AsO. 
The change in the bandwidth is due, in equal amount, to an increase of the band maximum and a decrease of the band minimum, with respect to the Fermi energy.
The band minimum goes from -2.15 to -3.2 eV, increasing the hybridization of Ru-4$d$ orbitals with the As-4$p$. 
The band maximum increases from 0.15 to  1.6 eV, showing that Ru related bands extend far beyond the Fermi level into the unoccupied states.  

\section{Abrikosov-Gor'kov theory}

In the main text we compare the $T_c$-suppression rates with standard Abrikosov-Gor'kov theory. 
In this section we briefly outline the procedure used to obtain those results. 

After averaging over random distributions of the impurities, the Green's function describing the electron recovers translational symmetry.
Thus, the full Green's function is generally given by
\begin{equation}\label{GAG}
 (G(\mathbf k,i\omega_n))^{-1}=(G^0(\mathbf k,i\omega_n))^{-1}-\Sigma(\mathbf k,i\omega_n),
\end{equation}
with 
\begin{equation}\label{G0AG}
 G^0(\mathbf k,i\omega_n)=(i \omega_n-H_0(\mathbf k)\rho_3- \Delta(\mathbf k) \rho_1 \sigma_2)^{-1},
\end{equation}
the Green's function in the impurity-free system, where $\omega_n=2\pi T(n+\frac{1}{2})$ and $T$ is the temperature. 
$\rho_i$ and $\sigma_i$ denote Pauli matrices operating on the electron and hole states and ordinary spin states, respectively. 
In the final spin-orbital-nambu space all quantities ($G^0(\mathbf k,i\omega_n)$, $\Sigma(\mathbf k,i\omega_n)$, and $G(\mathbf k,i\omega_n)$) are $10\times 10$ matrices.
The SC order parameter is obtained from the gap equation
\begin{equation}
 \Delta(\mathbf k)=\frac{T}{\mathcal{V}}\sum_{\omega_n,\mathbf k'} \Gamma(\mathbf k')\tr\left[\rho_1\sigma_2 G (\mathbf k',i\omega_n)\right],
\end{equation}
with $\Gamma(\mathbf k')=4\Gamma_{\mu}\cos k'_x \cos k'_y$ and in Born approximation the self-energy is given by
\begin{equation}\label{SelfAG}
 \Sigma(\mathbf k,i\omega_n)=\frac{n}{\mathcal{V}}\sum_{\mathbf k'} \Omega G (\mathbf k',i\omega_n) \Omega,
\end{equation}
where $\Omega=V_{\mu}\rho_3$ for nonmagnetic impurities and $\Omega=\sigma_3 IS_{\mu}\rho_0$ for magnetic impurities. 
The calculated Abrikosov-Gor'kov curves are obtained through iterative convergence of Eqs.~\eqref{GAG}-\eqref{SelfAG} using 600 Matsubara frequencies which was checked to be enough in the cases presented. 

We stress that interference between the impurities, and all the interaction effects discussed in the main text are neglected within Abrikosov-Gor'kov theory.

\section{Magnetic disorder in Sm-1111}

\begin{figure}[t!]
\begin{center}
 \includegraphics[width=0.99\textwidth]{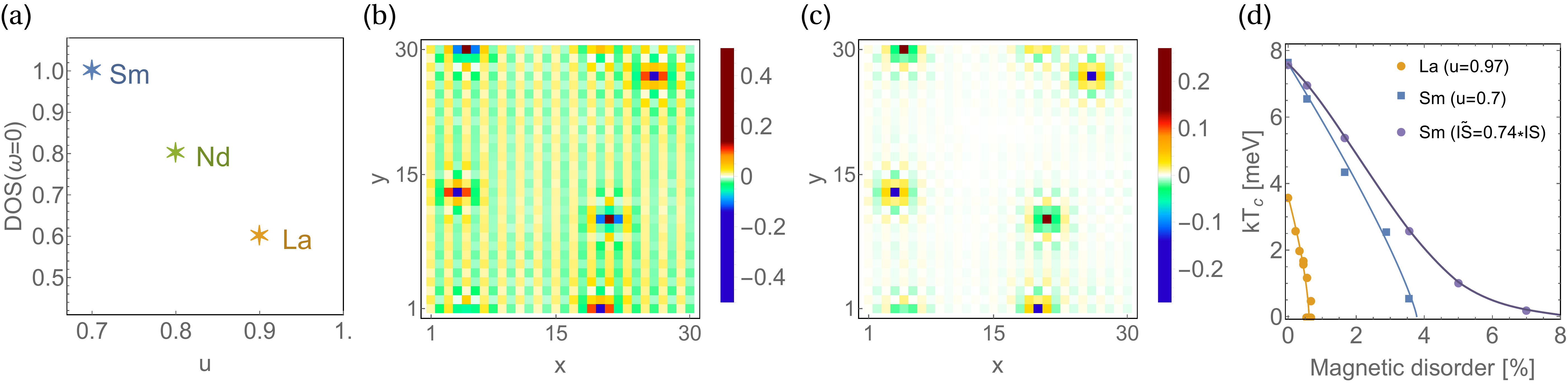}
\end{center}
    \caption{(a) Schematic phase diagram of the L-1111 (L=Sm, Nd, La) members as a function of correlations $u$ and DOS at the Fermi level. 
    Correlation-induced magnetic potential $I_{ind} s_{\mathbf{i}\mu}$ of the $d_{yz}$ orbital for (b) La-1111 $(u=0.97)$ and (c) Sm-1111 $(u=0.70)$ for a $0.55\%$ impurity configuration. 
    (d) $T_c$ as a function of magnetic impurity concentration for  La- and Sm-1111 systems. }
    \label{fig:Smmag}
\end{figure}

In this section we show the $T_c$-suppression as a function of magnetic impurity concentration for the Sm-1111 case, to be contrasted with the $T_c$-suppression of La-1111 (poisoning effect) presented in the manuscript. 
Figure~\ref{fig:Smmag}(a) shows a schematic phase diagram of the L-1111 (L=Sm, Nd, La) members within the present scenario. 
The main proposed differences between the compounds are:
1) Sm-1111 exhibits a larger DOS (than Nd- and La-11111) at the Fermi level (see $\gamma$ pocket evolution in Fig.~\ref{fig:sm1}) and hence a larger $T_c$ as explicitly shown in Fig.~\ref{fig:Smmag}(d),
and 2) Sm-1111 is less correlated (than Nd- and La-11111), and hence described by a $u$ parameter further away from the critical value $u_c$. 
The second assumption implies weaker correlation effects, i.e. the additional interaction-induced pair-breaking effect is diminished, as evident from comparison of Figs.~\ref{fig:Smmag}(b) ($u=0.97$ for La-1111) and \ref{fig:Smmag}(c) ($u=0.7$ for Sm-1111).
The long-range magnetic order is essentially gone in the less correlated system, and the local magnetic puddle is weaker as well.
As a consequence of the assumptions 1) and 2), it requires almost an order of magnitude larger magnetic impurity concentrations to destroy superconductivity in Sm-1111, as seen from Fig.~\ref{fig:Smmag}(d). 
In Fig.~\ref{fig:Smmag}(d), it is also seen that a $25\%$ weaker coupling of the Mn atoms with the conduction electrons, results in a $T_c$-suppression rate with an $8\%$ critical impurity concentration similar to that found experimentally in optimally doped Sm-1111~\cite{Ssingh11}.

\section{Comparison with $\mu$SR}

\begin{figure}[t!]
\begin{center}
 \includegraphics[width=0.6\textwidth]{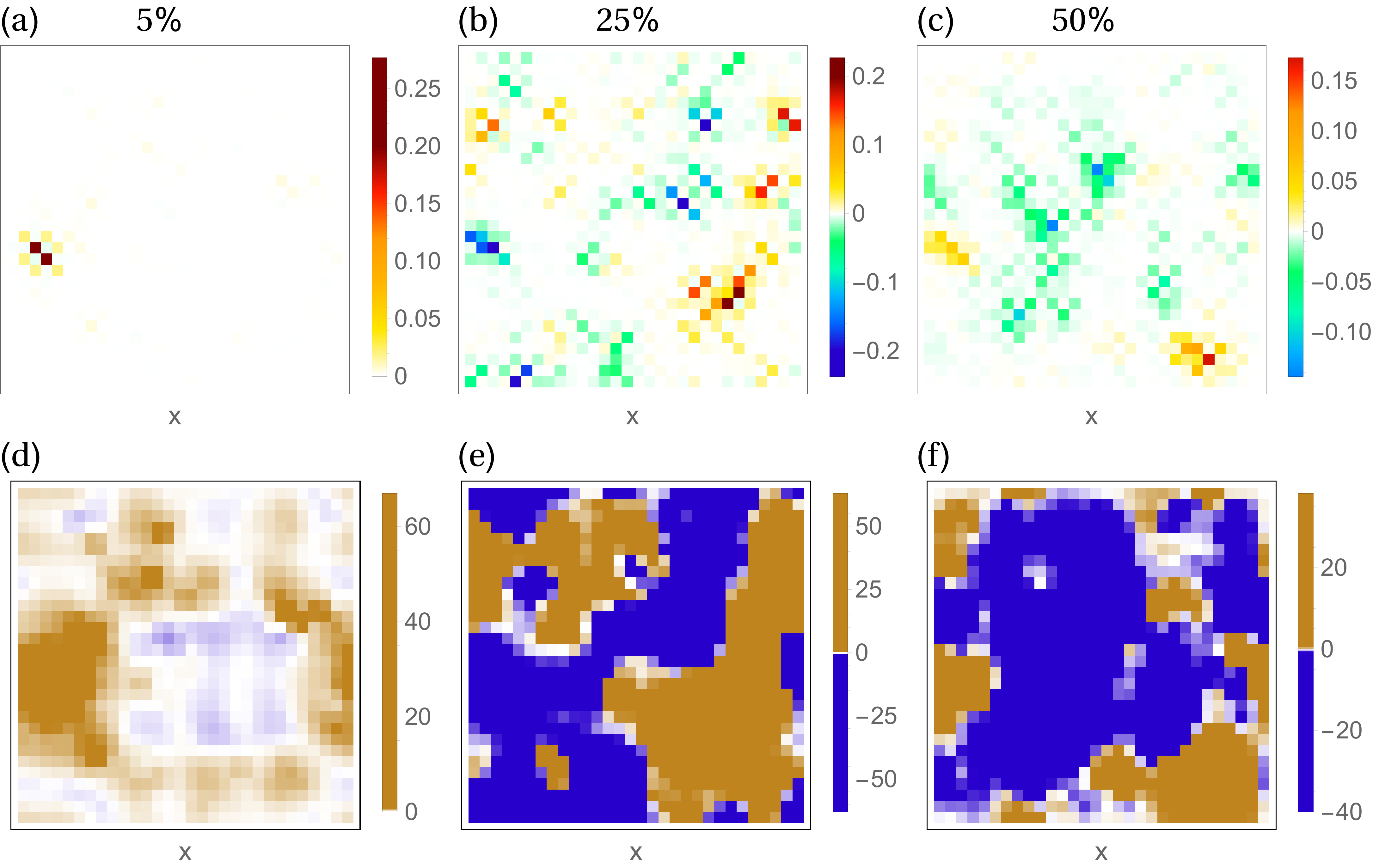}
\end{center}
    \caption{Induced magnetization $\mathbf{m}_\mathbf i$ with (a) 5\%, (b) 25\% and (c) 50\% non-magnetic disorder content. 
    Panels (d)-(f) display the respective fields $B(\mathbf{r})$ associated with (a)-(c). }
    \label{fig:muon2}
\end{figure}

The magnetic ordering temperature $T_m$ determined from $\mu$SR experiments is extracted from the $T$ evolution of the magnetic volume fraction defined by the fraction of muons that detect a local field exceeding $\sim 0.5$mT.\cite{SSannaPRB2010} 
Specifically, $T_m$ is defined as the highest $T$ where the volume fraction is $50\%$.
The local field is proportional to the staggered moment mainly through the dipolar coupling. 
Here we implement the map from a given staggered magnetization field at the muon sites $\mathbf{r}$:
\begin{equation}
 B(\mathbf{r})=\sum_{\mathbf{i}\mu} \frac{m_{\mathbf{i}\mu}}{|\mathbf{r}_\mathbf{i}|^3},
\end{equation}
where $\mathbf{r}_\mathbf{i}=\left(ax_\mathbf{i}-(a/2+a x),ay_\mathbf{i}-(a/2+a y),c_0\right)$ denotes the relative distance between the muon site $\mathbf{r}$ and the moment position at site $\mathbf{i}$. 
We have used the symmetric position of the main muon site $(a/2,a/2,c_0)$, with $a=2.83$\AA{} and $c_0=0.78$\AA{}~\cite{Smallett2}. 
We show examples of three magnetic states for different Ru concentration systems and their corresponding $B(\mathbf{r})$ in Fig.~\ref{fig:muon2}.

\begin{figure}
\begin{center}
 \includegraphics[width=0.7\textwidth]{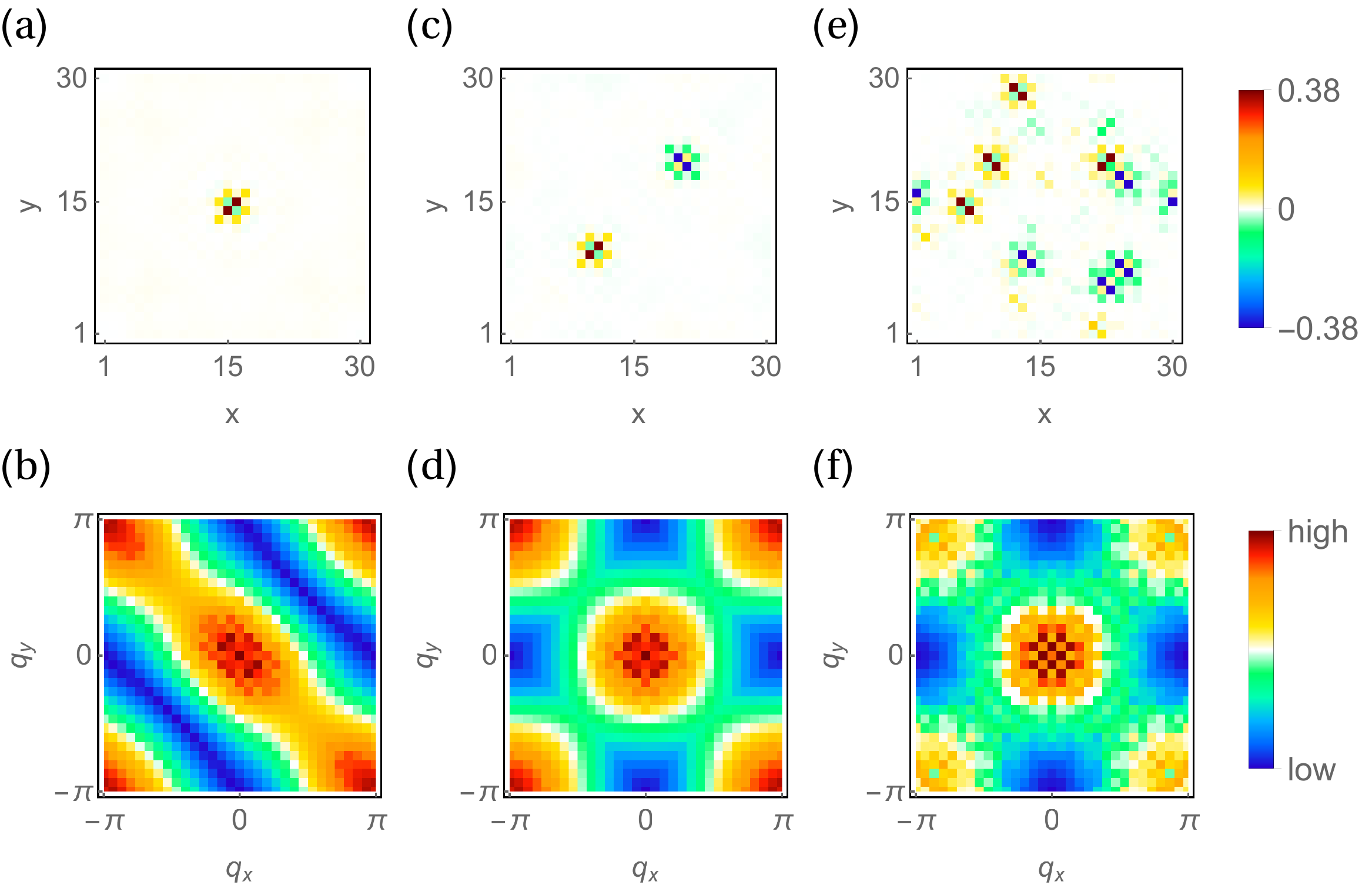}
\end{center}
    \caption{Dimer-induced magnetic structure. (a) Single dimer, (b) two dimers with opposite orientations, and (c) 15\% non-magnetic disorder concentration. 
    (d)-(f) The Fourier transformed $|m(\mathbf{q})|$ of the cases (a)-(c), respectively. 
    The 15\% case has been averaged over twelve different random configurations.}
    \label{fig:mq}
\end{figure}

Next, we turn to a brief discussion of the momentum structure of the magnetic phase induced by Ru substitution. 
The Fourier transformed magnetic structure of a single dimer shown Fig.~\ref{fig:mq}(a) can be seen in Fig.~\ref{fig:mq}(b). 
It consists of broad peaks at low wave-vectors and near $(\pi,\pi)$.
In general, adding Ru to the system will result in a roughly equal occupation of oppositely oriented dimer structures, as illustrated in Fig.~\ref{fig:mq}(c). 
Thus the overall structure factor in Fig.~\ref{fig:mq}(d) respects tetragonal symmetry. 
We verified that the magnetic structure of the disordered-induced magnetic phase remains dominated by the "single-dimer" results of Figs.~\ref{fig:mq}(a-d) by calculating the average of the magnetic structure factor for twelve different configurations with $15\%$ nonmagnetic disorder, shown in Fig.~\ref{fig:mq}(f).
The induced magnetic (short-range) order in Figs.~\ref{fig:mq}(e-f) is clearly very different from the stripe-like long-range magnetic order induced by magnetic impurities (see for example Fig.~\ref{fig:isinduced}(b)), with sharp $(\pi,0)/(0,\pi)$ peaks.

\section{Nonmagnetic disorder in La-1111}

Finally, we return briefly to the discussion of the distinction between the three different 1111 materials shown in Fig.~\ref{fig:Smmag}(a). A necessity for La-1111 to exhibit the poisoning effect in the case of magnetic disorder is the closeness to $u_c$ for this material as illustrated in Fig.~\ref{fig:Smmag}(a). For the case of nonmagnetic disorder, La-1111 will generally exhibit a pronounced magnetic phase, and in particular, exhibit a larger magnetic phase than e.g. Sm-1111 contrary to experimental findings. This, however, is only true for identical local Ru potentials in La-1111 and Sm-1111. Different material parameters will lead to differences in the relevant extracted Ru potentials. There is another effect, however, which becomes important for correlated systems, which is the additional screening caused by the Hubbard cost of charge modulations.~\cite{andersenthermal}

\begin{figure}[h]
\begin{center}
 \includegraphics[width=0.5\textwidth]{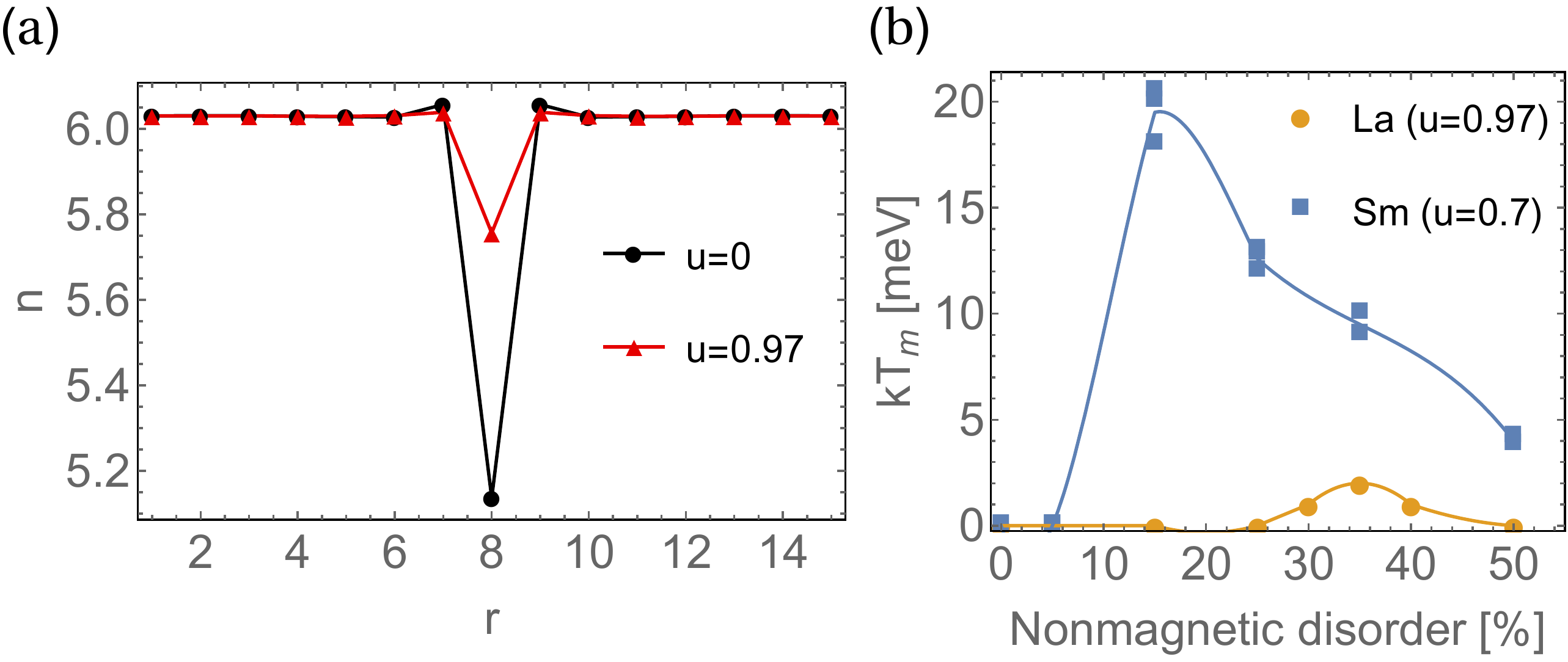}
\end{center}
    \caption{(a) Total electron density as a function of position along a cut through a single non-magnetic impurity. As seen by comparison of the black ($u=0.0$) and red curves ($u=0.97$) correlations weaken the charge modulations. (b) Comparison of Induced magnetic order for Sm-1111 (blue curve) and La-1111 (orange curve) showing a less pronounced magnetic phase for La-1111 despite its larger correlations.}
    \label{fig:Lanonmag}
\end{figure}

Figure~\ref{fig:Lanonmag}(a) shows the electron density along a line cut through a nonmagnetic potential. As seen, the correlations effectively screens the bare site potential. Such effects are typically not included in DFT-extracted impurity potentials, and could be an important source of discrepancy between bare and dressed potentials in correlated systems. In the current case, such renormalized potentials mainly relevant to La-1111 will shift the bound-state structure away from the Fermi level, locally weakening the Stoner condition and thereby compensating the larger $u$. 
We show in Fig.~\ref{fig:Lanonmag}(b) the magnetic dome for a renormalized $V_5=0.5$ eV in the correlated $u=0.97$ La-1111 system, much smaller than the dome of the Sm-1111 system despite the weaker correlations in the latter compound.

\end{document}